# Accurate quantum chemical free energies at affordable cost

GiovanniMaria Piccini[*,a],[b], Michele Parrinello[a],[b],[c]

[a]Department of Chemistry and Applied Biosciences, ETH Zurich, c/o USI Campus, Via Giuseppe Buffi 13, CH-6900, Lugano, Switzerland

[b]Facoltà di Informatica, Istituto di Scienze Computazionali, Università della SvizzeraItaliana (USI), Via Giuseppe Buffi 13, CH-6900, Lugano, Switzerland

[c]Istituto Italiano di Tecnologia, Via Morego 30, 16163 Genova, Italy

**Abstract**

Free energy sampling methods allow studying the full dynamics of activated processes. Unfortunately, the affordable accuracy of the potential describing the energy and forces of the system is usually rather low. Here we introduce a new method that combining metadynamics and free energy perturbation allows calculating accurate quantum chemical free energies for chemical reactions. To prove the effectiveness of this new approach we study the $S_N2$ reaction of $CH_3F + Cl^- \rightarrow CH_3Cl + F^-$ in vacuo and solvated by water. Comparisons are made with harmonic transition state theory to show how this method could provide accurate equilibrium and rate constants for complex systems.

Molecular dynamics (MD) simulations have become an indispensable tool in the study of a vast range of problems in chemistry. They allow a detailed atomistic description of several chemical processes bridging the gap between theory and experiments. A major interest in chemistry is the study of activated processes such as chemical reactions. These transformations imply the presence of high free energy barriers between the metastable states of interest, i.e. reactants and products. Transition state theory[1-3] (TST) tells us that the average transition time in an activated process scales exponentially with the associated free energy barrier. Unfortunately, for most of the reactions these time scales are incomparably larger than those that MD can reach. This fact limits dramatically the application of standard MD to the study of chemical reactions.

To overcome these kinetic bottlenecks different methods have been proposed starting from the landmark paper of Torrie and Valleau[4] in which umbrella sampling was introduced. Among many techniques and algorithms, metadynamics[5-8] (MetaD) has become one of the most popular and effective methods to enhance rare events sampling. In MetaD an adaptive bias potential is added to the underlying potential energy of the system to accelerate the transitions between metastable states. The bias is usually constructed iteratively during the simulation by adding a series of repulsive Gaussian kernels along collective variables (CVs), or order parameters that describe concisely the process of interest.

A good choice of the CVs is essential for the success of a MetaD simulation.[9] Chemical reactions can be seen as a rearrangement of the bonding topology of the molecular space. In such a complex scenario one could imagine that building CVs is a daunting task requiring either going to a multidimensional set of CVs[10] or using a complicated non-linear transformation of the atomic coordinates generalizing the concept of intrinsic reaction coordinate[11]. Unfortunately, in several cases these approaches are simply impracticable. Recently, we have proposed a simple yet elegant solution to this problem[12-13]. The method called harmonic linear discriminant analysis (HLDA), a modification of Fisher's classification theory of linear discriminants[14], requires as an input only the equilibration properties of the metastable states. The CVs derived in this scheme have shown to be very effective[15-17].

A central issue in quantum chemistry is accuracy[18-21]. In order to be predictive, a calculation should be able to reproduce experimental evidences not only qualitatively but also quantitatively. In principle free energy methods can provide the most accurate way to estimate equilibrium constants and reaction barriers. However, to converge the results millions of points in the configuration space must be sampled. Classical force fields would allow such sampling to be performed at a very low computational cost. However, their representation of the interatomic interactions is not meant to describe reactive events such as bond breaking and forming. Therefore, the use of quantum chemical methods is mandatory[22]. This limits dramatically the application of sampling techniques in terms of system size and complexity as well as the affordable accuracy of the quantum chemical based potential energy surface. Most of the times, this inaccurate description of the process allows drawing only qualitative conclusions and limiting predictability.

To circumvent this problem, we propose a method to obtain accurate estimates of reaction free energies and barriers by combining MetaD[8] with free energy perturbation[23-24] (FEP). We apply this new method to the calculation of the equilibrium properties and the estimation of rate constants of the $S_N2$ reaction $CH_3F + Cl^- \rightarrow CH_3Cl + F^-$. We demonstrate how starting from a low accuracy free energy surface (FES) obtained using the semi-empirical PM6[25] potential a higher level (DFT or MP2) quantum chemical FES can be obtained at a limited computational cost.

We start by running a MetaD simulation using a low-level method, e.g. a semi-empirical Hamiltonian[25] or density functional tight binding[26]. At convergence a restricted number of configurations is extracted from the trajectory, allowing a reasonably long simulation to be performed in a relatively limited time. Due to the presence of the history dependent bias potential every configuration carries a weight that reflects its statistical relevance and can be calculated from the MetaD theory[27-28]. These weights allows the reconstruction of the unbiased probability distribution and, therefore, of the FES. As recently suggested by Li et al.[29] by calculating the energy of such configurations at a different levels of theory it is possible to reconstruct the relative FES via free energy perturbation. The high-level FES is obtained through the relationship $F_{HL}(s) = F_{LL}(s) + \Delta F_{FEP}(s)$ where $F_{HL}$ and $F_{LL}$ are the high-level and low-level free energies calculated along a CV that is a function of the atomic Cartesian coordinates of the system $\mathbf{R}$: $s(R_1, R_2, R_3, ...)$, and where $\Delta F_{FEP}$ is the perturbation term. Given the potential energies for each configuration extracted from the original trajectory for both the low- and high-level Hamiltonian, $U_{LL}(\mathbf{R})$ and $U_{HL}(\mathbf{R})$, the perturbation term for a given value of the CV can be easily calculated as

$$\Delta F_{FEP}(s) = \frac{1}{\beta} \log \frac{\sum_i^N w_i^M(\mathbf{R}_i) \, w_i^P(\mathbf{R}_i) \delta(s-s(\mathbf{R}_i))}{\sum_i^N w_i^M(\mathbf{R}_i) \delta(s-s(\mathbf{R}_i))} \tag{1}$$

where $\beta$ is the inverse temperature $1/k_B T$, N is the number of configurations extracted from the low-level MetaD simulation, $w_i^M(\mathbf{R}_i)$ is the weight of the i-th sample extracted from MetaD and is calculated as $w_i^M(\mathbf{R}_i) = e^{\beta(V(s(\mathbf{R}_i),t)-c(t))}$ where $V(s(\mathbf{R}_i), t)$ is the value of the MetaD bias at for a given configuration $\mathbf{R}_i$ and time $t$, and $c(t)$ is an estimator of the reversible work done by the bias[27-28], and $w_i^P(\mathbf{R}_i)$ is the perturbative weight associated to the i-th sample that can be easily evaluated as $w_i^P(\mathbf{R}_i) = e^{\beta(U_{HL}(\mathbf{R}_i) - U_{LL}(\mathbf{R}_i))}$.

We studied the $S_N 2$ reaction using two different solvent models. In the first model ($M_1$) the presence of the solvent was implicit and its effect was felt by the reacting species in the form of an imposed thermostat. In the second model ($M_2$) the granularity of the solvent was fully taken into account using a QM/MM scheme in which the solvating water molecules were modeled as classical particles. The chemical process was initially described using the PM6 semi-empirical model. Technical aspects of the simulation are reported in detail in the SI. Here we sketch only the main steps of the procedure.

In order to apply HLDA and derive optimal CVs we first need to assign descriptors that are able to distinguish between the two metastable states. A natural choice is to use (see Fig 1) the distances $d_1$ and $d_2$ of the two halides from the carbon atom. Following the HLDA prescription we calculate in two different unbiased runs the average value and the fluctuation matrix of the descriptors in the two metastable states. This information was combined as in refs.[12-13] to give the following CV:

$s = 0.81 \cdot d_1 - 0.59 \cdot d_2$ for $M_1$

and

$s = 0.90 \cdot d_1 - 0.43 \cdot d_2$ for $M_2$

In both cases already at this stage the HLDA derived CVs are capable of capturing the chemistry of the process that results from the antisymmetric[30] displacement of the two halides in which the $d_1$ has more weight due to the larger strength of the C-F bond.

Well-tempered metadynamics[6] runs were performed on the two systems using the corresponding CVs. These runs lead to the free energies depicted in Fig 2, that exhibit two minima corresponding to reactants and products.

Metadynamics trajectories 1 ns long were run for the two models. After discarding the initial equilibration segment, we extracted from the trajectories about 2000 configuration, which amounts a considering only 1 out of every 1000 configuration generated.

For both systems FEP corrections to the free energy surface have been calculated using two methods of increasing accuracy as well as computational cost, a DFT GGA potential such as PBE[31] with Grimme's semi-empirical D3 dispersion correction[32-33] and a wavefunction-based correlated method such as MP2[34] (further computational details are available in the SI). Whereas nowadays full ab initio simulations using DFT potentials have become more affordable due to the increase in computer power and gain in algorithmic efficiency [35], the same is not yet possible for expensive correlated post-Hartree-Fock methods[36]. In contrast to this, MetaFEP needs only relatively few configurations and only relative energy difference estimations to converge the high-level free energy as reported in the results section.

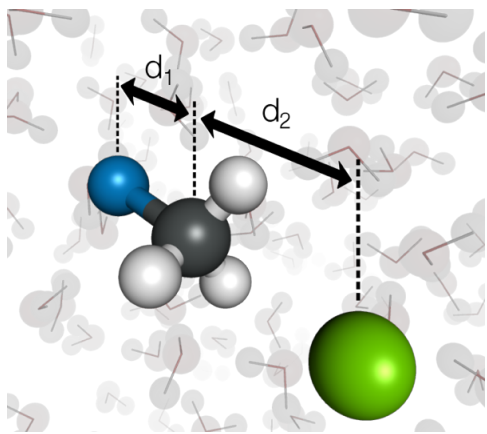

*Figure 1 Essential carbon-halide distances describing the $S_N2$ reaction used as basis for the HLDA approach.*

First, we discuss the results for the model $M_1$. We calculated the FEP term using eq.(1). This quantity added to the low-level FES provides an accurate estimation of the free energy of the system at a higher level of theory. It is important to point out that for this type of calculations *no gradients* are needed as the perturbation depends only on the energy difference of the configurations and no dynamics has to be propagated. Furthermore, this part of the calculation is trivially parallelizable. This is a further great advantage in terms of computational cost as the calculation of the interatomic forces can be extremely expensive. Since the term of eq.(1) could be noisy in some regions of the CV space (see supporting material for raw data), to smooth the results we applied a moving average filter to the data. In such way we can also calculate the standard deviation of the perturbative term, thus, allowing the calculation of confidence intervals. Figure 2 reports the MetaFEP results for the reaction in vacuo using PBE+D and MP2. It is evident how the present method allows estimating the free energy profile even when the potential energy of the system at the two levels of theory differs substantially.

In this case the free energy differences obtained with MetaFEP could be conveniently compared to the those obtained by calculating the free energy of the reference states using the harmonic approximation (HA) to include the vibrational thermal contributions[37] according to TST[1].Reference states for the harmonic analysis were obtained by optimizing reactants, products and transition structures at 0 K. Table 1 compares the reaction and activation free energies for model $M_1$ calculated by integrating the free energy basins within the reactants, products and transition state domains for the MetaFEP free energy surfaces and by applying harmonic TST respectively. The accuracy of MetaFEP is surprisingly high, especially for the reaction free energies. Relatively larger discrepancies are observed when comparing the activation energies. Two main sources of error can be associated to this effect. On the one hand, enhanced sampling methods such as MetaD do not sample accurately the transition region. Thus, larger uncertainties are always expected when extracting activation energies whereas more specialized variants of the method like infrequent MetaD[38] or variational flooding[30, 39-40] provide a more formal theoretical framework to extract this information. On the other hand, harmonic TST accuracy is strongly limited by the rough approximation made by assuming the normal modes of vibration to be perfectly harmonic. This introduces large inaccuracies especially in the description of the transition structure that may be strongly affected by anharmonic effects[3, 41]. Having in mind the possible errors when estimating the activation free energies from both approaches, we can state that the accuracy of the present approach is more than satisfactory even for the calculation of reaction free energy barriers.

Table 1 Reaction, $\Delta F_{B-A}$, and activation, $\Delta F^{\ddagger}_A$ and $\Delta F^{\ddagger}_B$, free energies for reaction model $M_1$ calculated using free energy perturbation with metadynamics, MetaFEP, and including harmonic vibrational thermal contributions at the PBE+D and MP2 levels of theory. Energies are expressed in kJ mol$^{-1}$.

|  | PBE+D | | MP2 | |
| --- | --- | --- | --- | --- |
|  | MetaFEP | HA | MetaFEP | HA |
| $\Delta F_{B-A}$ | 85 | 85 | 98 | 98 |
| $\Delta F^{\ddagger}_A$ | 94 | 88 | 126 | 120 |
| $\Delta F^{\ddagger}_B$ | 8 | 2 | 28 | 22 |

To prove the effectiveness and wide applicability of the present method, as well as its simplicity, in treating complex chemical systems we studied this reaction in explicit water applying a QM/MM approach (model $M_2$). To be consistent with the results obtained for model $M_1$ and to be able to qualitatively compare the change in free energy barriers and relative stability we used the same levels of theory, namely PM6 for MetaD, and PBE+D and MP2 for MetaFEP, to describe the QM region of the system, while the TIP3P potential has been used consistently to represent the classical interaction between water molecules and the QM region within a mechanical embedding scheme[42]. Fig. 3 reports these results. It is observed that the lowering of the barrier and relative stability of reactants and products follow the same trend of the reaction model $M_1$ with differences that are consistent with the changes observed in the PM6 MetaD simulation. In this case calculation of harmonic frequencies to approximate the free energy differences would be meaningless and a direct comparison with other free energy methods would be computationally prohibitive.

We conclude that this new method allows estimating chemically accurate equilibrium constants and nearly chemically accurate free energy barriers and henceforth rate constants for complex chemical reactions. The practical possibility of stepping up the accuracy of the calculation greatly extends the scope and reliability of metadynamics. In some cases, the low-level description of the dynamics may not be sufficiently adequate to explore the most representative high-level configurations. In these situations, more advance techniques allowing sampling of specific regions of the phase space will be necessary such as variational enhanced sampling[39]. These cases, however, represent a relatively small class of problems in chemistry for which a proper understanding of the electronic structure effects on the reaction mechanism allows to spot the main criticalities and to prevent inaccurate results.

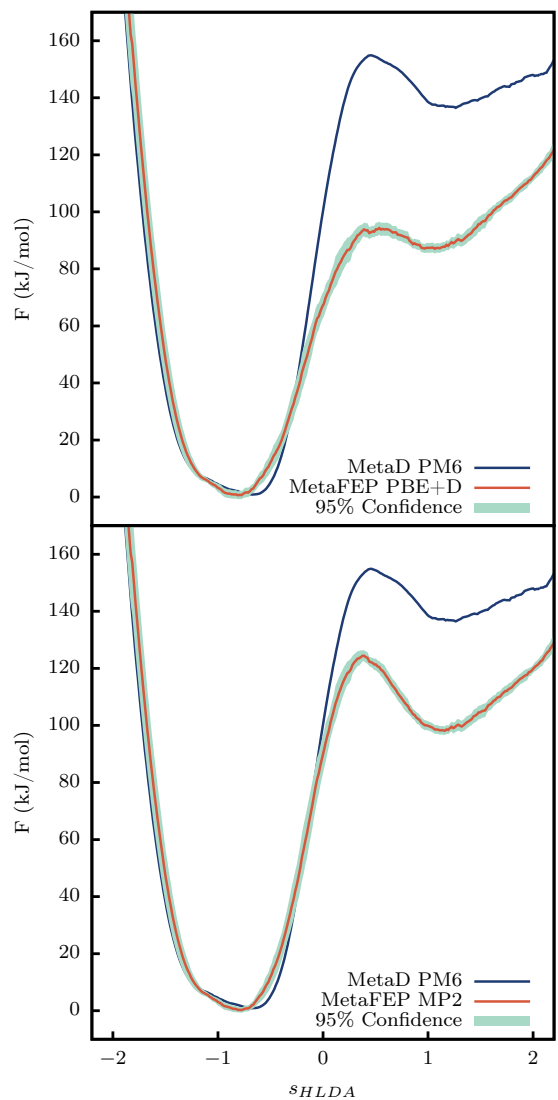

*Figure 2 Free energy along the HLDA CV calculated using MetaD with the semi-empirical PM6 Hamiltonian and using MetaFEP for the S$_N$2 reaction model M$_1$. MetaFEP results are smoothed using a moving average over data points providing relative confidence intervals. Upper panel: FEP results for the PBE+D DFT GGA potential with D3 semi-empirical dispersion. Lower panel: FEP results for the MP2 wavefunction-based correlation method.*

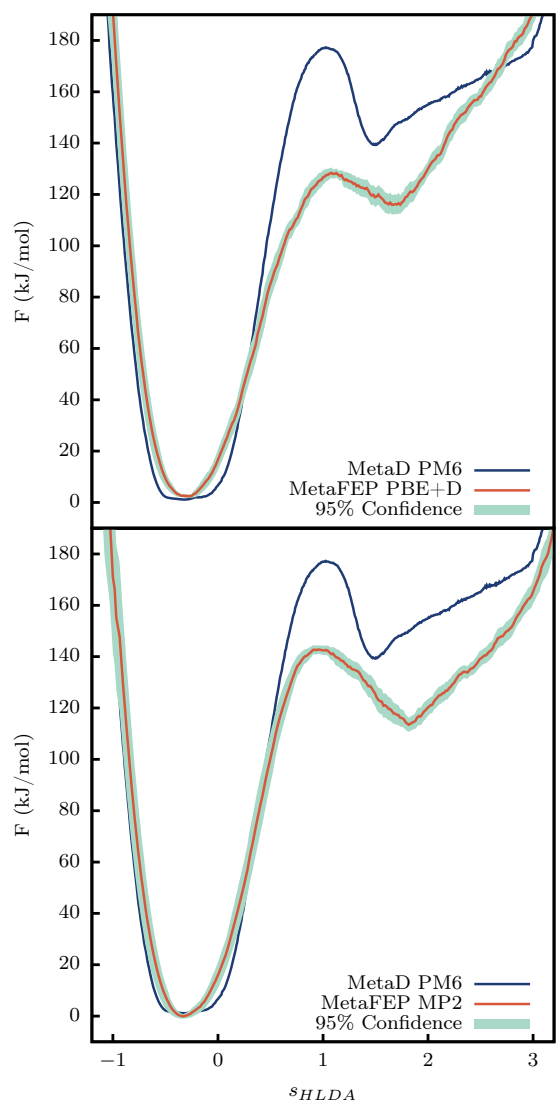

*Figure 3 Free energy along the HLDA CV calculated using MetaD with the semi-empirical PM6 Hamiltonian and using MetaFEP for the $S_N2$ reaction model $M_2$. MetaFEP results are smoothed using a moving average over data points providing relative confidence intervals. Upper panel: FEP results for the PBE+D DFT GGA potential with D3 semi-empirical dispersion. Lower panel: FEP results for the MP2 wavefunction-based correlation method.*

**References**

[1] H. Eyring, *J. Chem. Phys.* **1935**, *3*, 107-115.
[2] B. Peters, B. Peters, in *Reaction Rate Theory and Rare Events Simulations*, **2017**, pp. 227-271.
[3] D. G. Truhlar, B. C. Garrett, S. J. Klippenstein, *J. Phys. Chem.* **1996**, *100*, 12771-12800.
[4] G. M. Torrie, J. P. Valleau, *J. Comput. Phys.* **1977**, *23*, 187-199.
[5] A. Laio, M. Parrinello, *Proc. Natl. Acad. Sci. U.S.A.* **2002**, *99*, 12562-12566.
[6] A. Barducci, G. Bussi, M. Parrinello, *Phys. Rev. Lett.* **2008**, *100*.
[7] A. Barducci, M. Bonomi, M. Parrinello, *WIRES: Comput. Mol. Sci.* **2011**, *1*, 826-843.
[8] O. Valsson, P. Tiwary, M. Parrinello, *Annu. Rev. Phys. Chem.* **2016**, *67*, 159-184.
[9] G. Piccini, D. Polino, M. Parrinello, *J. Phys. Chem. Lett.* **2017**, *8*, 4197-4200.
[10] F. Pietrucci, W. Andreoni, *Phys. Rev. Lett.* **2011**, *107*, 085504.
[11] C. Gonzalez, H. B. Schlegel, *J. Chem. Phys.* **1989**, *90*, 2154-2161.
[12] D. Mendels, G. Piccini, M. Parrinello, *J. Phys. Chem. Lett.*, *0*, 2776-2781.
[13] G. Piccini, D. Mendels, M. Parrinello, *J. Chem. Theory and Comput.* **2018**, *14*, 5040-5044.
[14] R. A. Fisher, *Annals of Eugenics* **1936**, *7*, 179-188.
[15] R. Capelli, A. Bochicchio, G. Piccini, R. Casasnovas, P. Carloni, M. Parrinello, *bioRxiv* **2019**, 544577.
[16] Y.-Y. Zhang, H. Niu, G. Piccini, D. Mendels, M. Parrinello, *J. Chem. Phys.* **2019**, *150*, 094509.
[17] D. Mendels, G. Piccini, Z. F. Brotzakis, Y. I. Yang, M. Parrinello, *J. Chem. Phys.* **2018**, *149*, 194113.
[18] T. Helgaker, T. A. Ruden, P. Jørgensen, J. Olsen, W. Klopper, *J. Phys. Org. Chem.* **2004**, *17*, 913-933.
[19] F. Claeyssens, J. N. Harvey, F. R. Manby, R. A. Mata, A. J. Mulholland, K. E. Ranaghan, M. Schütz, S. Thiel, W. Thiel, H. J. Werner, *Angew. Chem. Int. Ed.* **2006**, *118*, 7010-7013.
[20] G. Piccini, M. Alessio, J. Sauer, *Angew. Chem. Int. Ed.* **2016**, *55*, 5235-5237.
[21] T. Schwabe, S. Grimme, *Phys. Chem. chem. Phys.* **2006**, *8*, 4398-4401.
[22] R. Car, M. Parrinello, *Phys. Rev. Lett.* **1985**, *55*, 2471-2474.
[23] C. Chipot, A. Pohorille, *Free energy calculations*, Springer, **2007**.
[24] R. W. Zwanzig, *J. Chem. Phys.* **1954**, *22*, 1420-1426.
[25] J. J. P. Stewart, *J. Mol. Model.* **2007**, *13*, 1173-1213.
[26] W. M. C. Foulkes, R. Haydock, *J. Phys. Rev. B* **1989**, *39*, 12520.
[27] M. Bonomi, A. Barducci, M. Parrinello, *J. Comput. Chem.* **2009**, *30*, 1615-1621.
[28] P. Tiwary, M. Parrinello, *J. Phys. Chem. B* **2015**, *119*, 736-742.
[29] P. Li, X. Jia, X. Pan, Y. Shao, Y. Mei, *J. Chem. Theory and Comput.* **2018**, *14*, 5583-5596.
[30] G. Piccini, J. J. McCarty, O. Valsson, M. Parrinello, *J. Phys. Chem. Lett.* **2017**, *8*, 580-583.
[31] J. P. Perdew., K. Burken, M. Ernzerhof, *Phys. Rev. Lett.* **1996**, *77*, 3865-3868.
[32] S. Grimme, J. Antony, S. Ehrlich, H. Krieg, *J. Chem. Phys.* **2010**, *132*, 154104.
[33] S. Grimme, S. Ehrlich, L. Goerigk, *J. Comput. Chem.* **2011**, *32*, 1456-1465.
[34] S. Kossmann, F. Neese, *J. Chem. Theory and Comput.* **2010**, *6*, 2325-2338.
[35] D. Marx, J. Hutter, *Ab initio molecular dynamics: basic theory and advanced methods*, Cambridge University Press, **2009**.

[36]   M. Del Ben, J. Hutter, J. VandeVondele, *J. Chem. Phys.* **2015**, *143*, 102803.
[37]   C. J. Cramer, F. Bickelhaupt, *Angew. Chem. Int. Ed.* **2003**, *42*, 381-381.
[38]   P. Tiwary, M. Parrinello, *Phys. Rev. Lett.* **2013**, *111*, 230602.
[39]   J. Debnath, M. Invernizzi, M. Parrinello, *arXiv preprint arXiv:.09032* **2018**.
[40]   J. McCarty, O. Valsson, P. Tiwary, M. Parrinello, *Phys. Rev. Lett.* **2015**, *115*, 70601.
[41]   D. G. Truhlar, B. C. Garrett, *Annu. Rev. Phys. Chem.* **1984**, *35*, 159-189.
[42]   A. W. Götz, M. A. Clark, R. C. Walker, *J. Comput. Chem.* **2014**, *35*, 95-108.

# Supporting information to: "Accurate quantum chemical free energies at affordable cost"

GiovanniMaria Piccini[*,**], Michele Parrinello[*,**,***]

*Department of Chemistry and Applied Biosciences, ETH Zurich, c/o USI Campus, Via Giuseppe Buffi 13, CH-6900, Lugano, Switzerland

**Facoltà di Informatica, Istituto di Scienze Computazionali, Università della SvizzeraItaliana (USI), Via Giuseppe Buffi 13, CH-6900, Lugano, Switzerland

***Istituto Italiano di Tecnologia, Via Morego 30, 16163 Genova, Italy

## Computational details model M$_1$

## PM6 metadynamics:

Energy and forces
- PM6 Hamiltonian
- SCF convergence 1.0E-5 hartree

Molecular dynamics
- total simulation time: 1 ns
- timestep: 0.5 fs
- temperature: 300 K
- thermostat: Bussi velocity resscaling[1]
- thermostat coupling frequency: 100 steps

Well-tempered metadynamics
- Gaussian hills height: 2.0 kJ/mol
- Gaussian hills sigma: 0.1 Å
- deposition rate (pace): 50 steps
- biasfactor γ: 50
- distance $d_1$ restraint: applied above +4 Å with k=150 kJ/mol Å$^2$
- distance $d_2$ restraint: applied above +4 Å with k=150 kJ/mol Å$^2$

CODES: CP2K[2]+PLUMED2[3]

## PBE+D energy evaluation

Energy and forces:

- SCF convergence 1.0E-8 hartree
- basis set: ma-def2-TZVPP (minimally augmented diffuse def2 basis set)
- Grimme's D3[4] dispersion with Becke-Johnson damping[5]

CODE: ORCA 4.0.0[6]

## MP2 energy evaluation

Energy and forces:

- SCF convergence 1.0E-8 hartree
- RI-MP2
- basis set: aug-cc-pVTZ (diffuse cc-pVTZ basis sets)
- auxiliary basis set: aug-cc-pVTZ/C
- RI-JK approximation for Coulomb and Exchange integrals
- RI-JK auxiliary basis set: def2/JK

CODE: ORCA 4.0.0[6]

## Computational details model $M_1$

## QM/MM PM6/TIP3P metadynamics:

Energy and forces

- PM6 Hamiltonian + TIP3P water model
- SCF convergence 1.0E-8 kcal/mol
- mechanical embedding
- truncated real space Ewald sum at 8 Å

Molecular dynamics

- total simulation time: 1 ns
- timestep: 0.5 fs
- temperature: 300 K
- thermostat: Bussi velocity resscaling[1]

Well-tempered metadynamics

- Gaussian hills height: 2.0 kJ/mol
- Gaussian hills sigma: 0.1 Å
- deposition rate (pace): 50 steps
- biasfactor γ: 50
- distance $d_1$ restraint: applied above +4 Å with k=150 kJ/mol Å$^2$
- distance $d_2$ restraint: applied above +4 Å with k=150 kJ/mol Å$^2$

CODES: AMBER16[7]+PLUMED2[3]

## PBE+D energy evaluation

Energy and forces:

- SCF convergence 1.0E-8 hartree
- basis set: ma-def2-TZVPP (minimally augmented diffuse def2 basis set)
- Grimme's D3[4] dispersion with Becke-Johnson damping[5]

CODE: ORCA 4.0.0[6]

## MP2 energy evaluation

Energy and forces:

- SCF convergence 1.0E-8 hartree
- RI-MP2
- basis set: aug-cc-pVTZ (diffuse cc-pVTZ basis sets)
- auxiliary basis set: aug-cc-pVTZ/C
- RI-JK approximation for Coulomb and Exchange integrals
- RI-JK auxiliary basis set: def2/JK

CODE: ORCA 4.0.0[6]

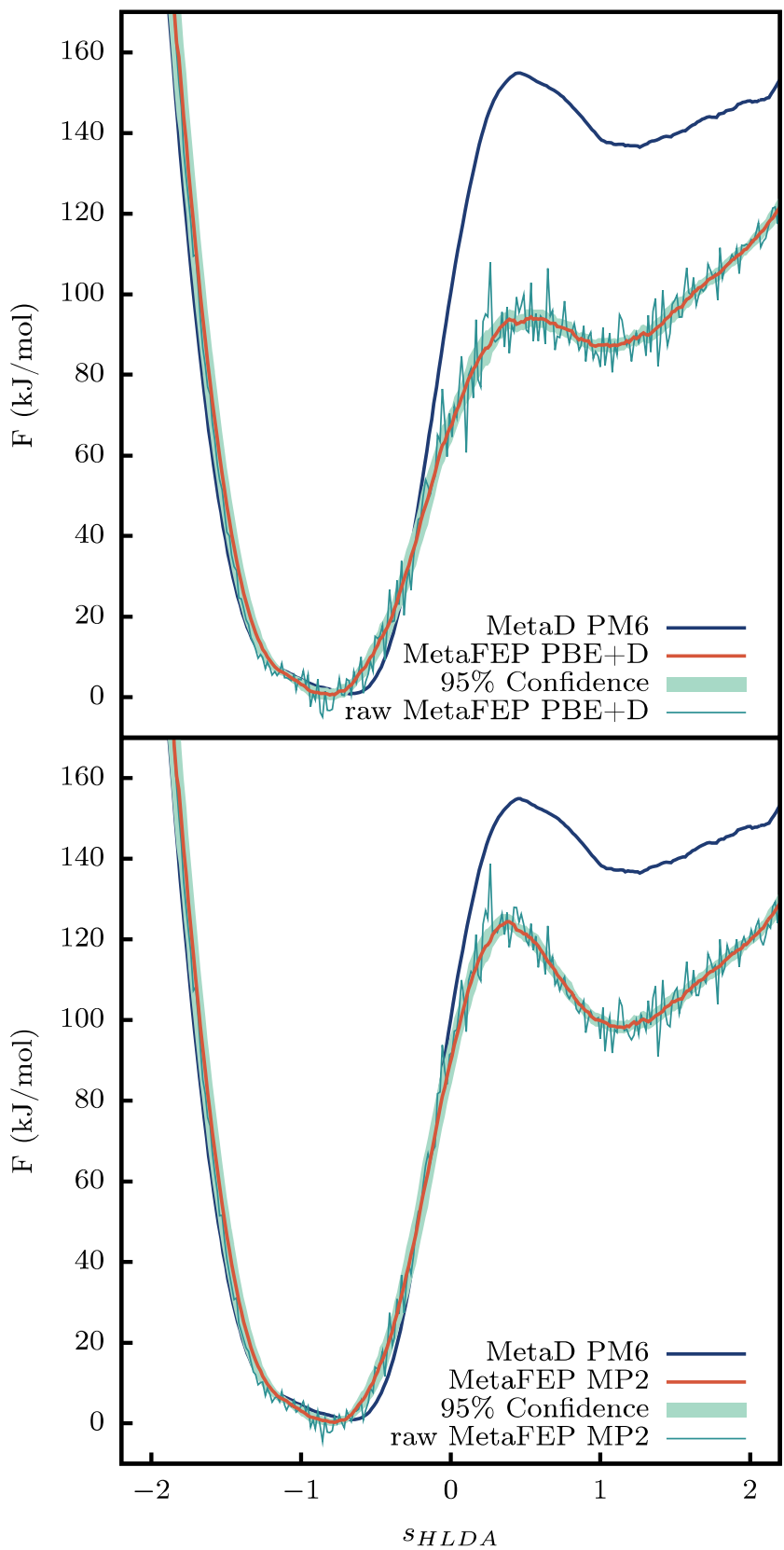

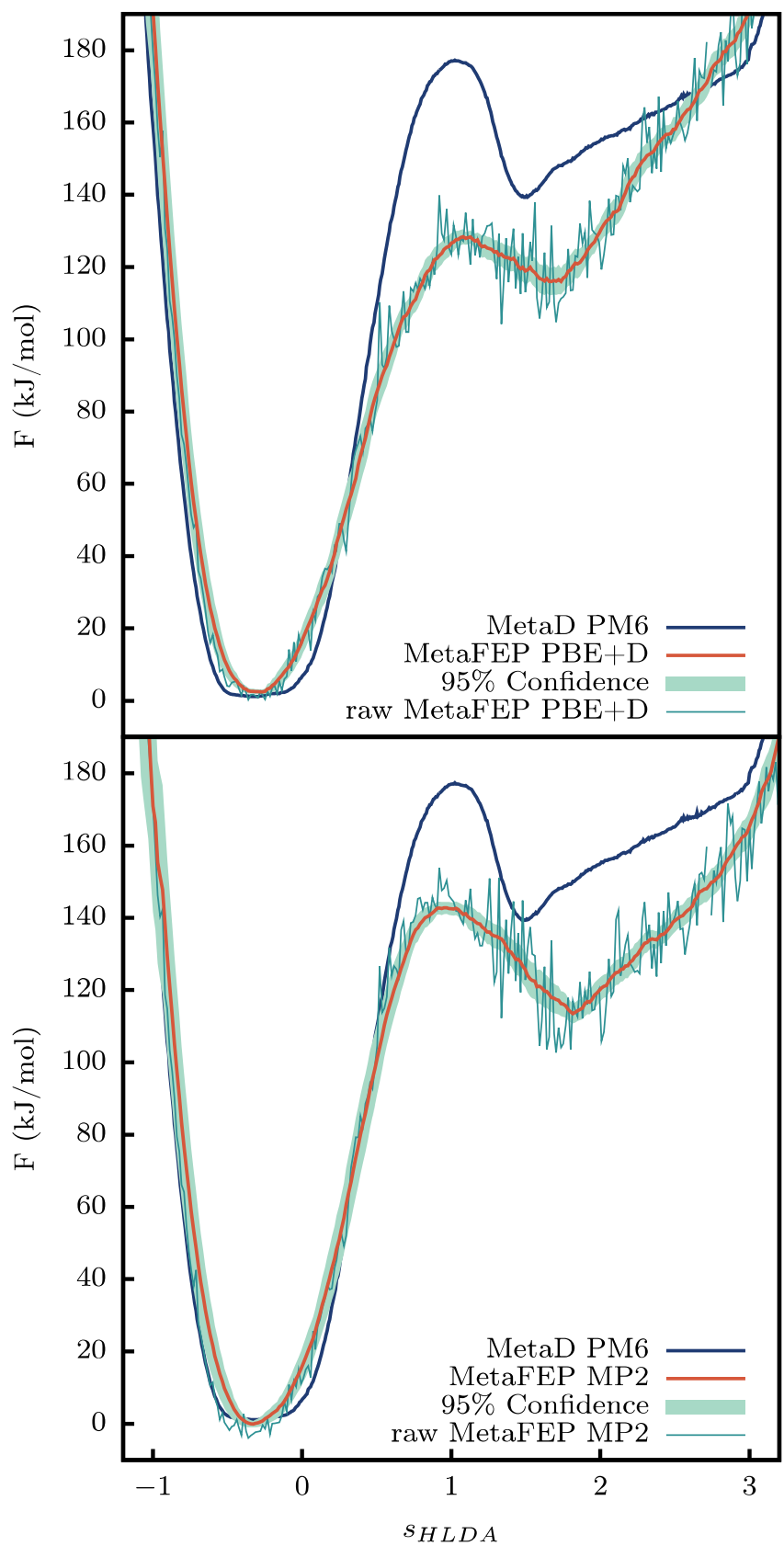

# References

[1] G. Bussi, D. Donadio, M. Parrinello, *J. Chem. Phys.* **2007**, *126*, 14101.
[2] J. Hutter, M. Iannuzzi, F. Schiffmann, J. VandeVondele, *WIREs Comput Mol Sci* **2014**, *4*, 15-25.
[3] G. A. Tribello, M. Bonomi, D. Branduardi, C. Camilloni, G. Bussi, *Computer Physics Communications* **2014**, *185*, 604-613.
[4] S. Grimme, J. Antony, S. Ehrlich, H. Krieg, *J. Chem. Phys.* **2010**, *132*, 154104.
[5] S. Grimme, S. Ehrlich, L. Goerigk, *J. Comput. Chem.* **2011**, *32*, 1456-1465.
[6] F. Neese, *Wiley Interdiscip Rev Comput Mol Sci* **2018**, *8*, e1327.
[7] aD. A. Case, T. E. Cheatham, T. Darden, H. Gohlke, R. Luo, K. M. Merz, A. Onufriev, C. Simmerling, B. Wang, R. J. Woods, *J. Comput. Chem.* **2005**, *26*, 1668-1688; bA. W. Götz, M. A. Clark, R. C. Walker, *J. Comput. Chem.* **2014**, *35*, 95-108; cR. C. Walker, M. F. Crowley, D. A. Case, *J. Comput. Chem.* **2008**, *29*, 1019-1031.